\documentclass[12pt]{iopart}
\usepackage{iopams}
\usepackage{graphicx} 
\usepackage{subfigure}
\usepackage{epstopdf}
\usepackage{lineno}
\usepackage{multirow}
\usepackage{rotating}
\usepackage{xspace}
\usepackage{cite,./mcite}
\usepackage{hyperref}

\usepackage{amssymb,amssymb}
\usepackage{amsthm}

\newlength{\figwidth}
\def\Wboson{\ensuremath{W}}%
\def\Zboson{\ensuremath{Z}}
\def\antibar#1{\ensuremath{#1\bar{#1}}}

\def\Zll{\ensuremath{Z \rightarrow \ell\ell}}
\def\Wln{\ensuremath{W \rightarrow \ell\nu}}
\def\Wen{\ensuremath{W \rightarrow e\nu}}
\def\Wmn{\ensuremath{W \rightarrow \mu\nu}}
\def\ttbar{\antibar{t}}
\def\met{\ensuremath{E_{\mathrm{T}}^{\mathrm{miss}}}} 
\def\tbar{\ensuremath{\bar{t}}}
\def\ifb{\mbox{fb$^{-1}$}}
\def\GeV{\ifmmode {\mathrm{\ Ge\kern -0.1em V}}\else
                   \textrm{Ge\kern -0.1em V}\fi}%
\def\MeV{\ifmmode {\mathrm{\ Me\kern -0.1em V}}\else
                   \textrm{Me\kern -0.1em V}\fi}%
\let\gev=\GeV
\let\mev=\MeV
\def\Wen{\ensuremath{W \rightarrow e\nu}}
\def\Wmn{\ensuremath{W \rightarrow \mu\nu}}
\def\pt{\ensuremath{p_{\mathrm{T}}}\xspace} 

\usepackage{preprintcover}  
\PreprintCoverPaperTitle{Search for $\Wboson H$ production with a light Higgs boson decaying to prompt electron-jets in proton--proton collisions at $\sqrt{s}=7$ TeV with the ATLAS detector}  
\PreprintIdNumber{CERN-PH-EP-2012-365}  
\PreprintCoverAbstract{A search is performed for $\Wboson H$ production with a light Higgs boson decaying to hidden-sector particles resulting in clusters of collimated electrons, known as electron-jets. The search is performed with 2.04 $\ifb$ of data collected in 2011 with the ATLAS detector at the LHC in proton--proton collisions at $\sqrt{s}=7$ TeV. 
~One event satisfying the signal selection criteria is observed, which is consistent with the expected background rate.
 Limits on the product of the  $\Wboson H$ production cross section and the branching ratio of a Higgs boson decaying to prompt electron-jets are calculated as a function of a Higgs boson mass in  the range from 100 GeV to 140 GeV.}  
\PreprintJournalName{New.~J.~Phys.}  

\begin{document}
\vspace*{-0.5in}


\title[]{Search for $\Wboson H$ production with a light Higgs boson decaying to prompt electron-jets in proton--proton collisions at $\sqrt{s}=7$ TeV with the ATLAS detector}

\author{The ATLAS Collaboration}


\date{18 March 2011}


\begin{abstract}
A search is performed for $\Wboson H$ production with a light Higgs boson decaying to hidden-sector particles resulting in clusters of collimated electrons, known as electron-jets. The search is performed with 2.04 $\ifb$ of data collected in 2011 with the ATLAS detector at the LHC in proton--proton collisions at $\sqrt{s}=7$ TeV. 
~One event satisfying the signal selection criteria is observed, which is consistent with the expected background rate.
 Limits on the product of the  $\Wboson H$ production cross section and the branching ratio of a Higgs boson decaying to prompt electron-jets are calculated as a function of a Higgs boson mass in  the range from 100 GeV to 140 GeV.
\end{abstract}


\maketitle


\section{Introduction}
Recently, the production of a boson with a mass of about 125 GeV has been observed by the ATLAS \cite{higgs_discovery} and CMS \cite{cms_higgs} collaborations. The observation is compatible with the production and decay of the Standard Model (SM) Higgs boson \cite{higgs1,higgs2,higgs3} at this mass. 
Strengthening or rejecting the SM Higgs boson hypothesis is currently of utmost importance and thus a search for non-SM Higgs boson decays is of high interest. In this Letter, a search for a Higgs boson decaying to a new hidden sector of particles is presented.
The masses and couplings of the hidden-sector particles are chosen such that the Higgs boson decay cascade results in jets consisting exclusively of electrons (``electron-jets'') and weakly interacting neutral particles  \cite{hidden_higgs_decays,discovering_hidden_higgs}. This is the first search performed for this particular channel. Moreover, in addition to a recently discovered state consistent with the SM Higgs boson, there may be other scalar fields that couple to the $\Wboson$ boson and decay to electron-jets. These scalars arise in the Higgs boson sector in many extensions of the SM, and electron-jets could be the primary discovery channel for these new states. 
The search is performed in the Higgs boson mass range between 100 GeV and 140 GeV. 
The analysis examines the associated Higgs boson production mechanism, $pp \to WH$, assuming SM couplings between the Higgs boson and the $\Wboson$ boson.

Many models of physics beyond the SM contain a light hidden sector, 
which is composed of as yet unobserved fields that are singlets under the SM group $SU(3)\times SU(2)\times U(1)$ and that can be probed at the LHC \cite{hidden_higgs_decays,discovering_hidden_higgs, Strassler2007374, nima_dark_matter_theory,nima_08100714,Baumgart_0901028, higgs_portal, Davoudiasl2005117, Gross1985253, PhysRevLett_98_221601, Holdom198665, Dienes_9610479v2}. Models of this hidden sector vary from simple modifications of the SM \cite{higgs_portal,Davoudiasl2005117} to models motivated by string theory \cite{Gross1985253}  to so-called unparticle models \cite{PhysRevLett_98_221601}.

\begin{figure}
  \centering
    \includegraphics[scale=0.28]{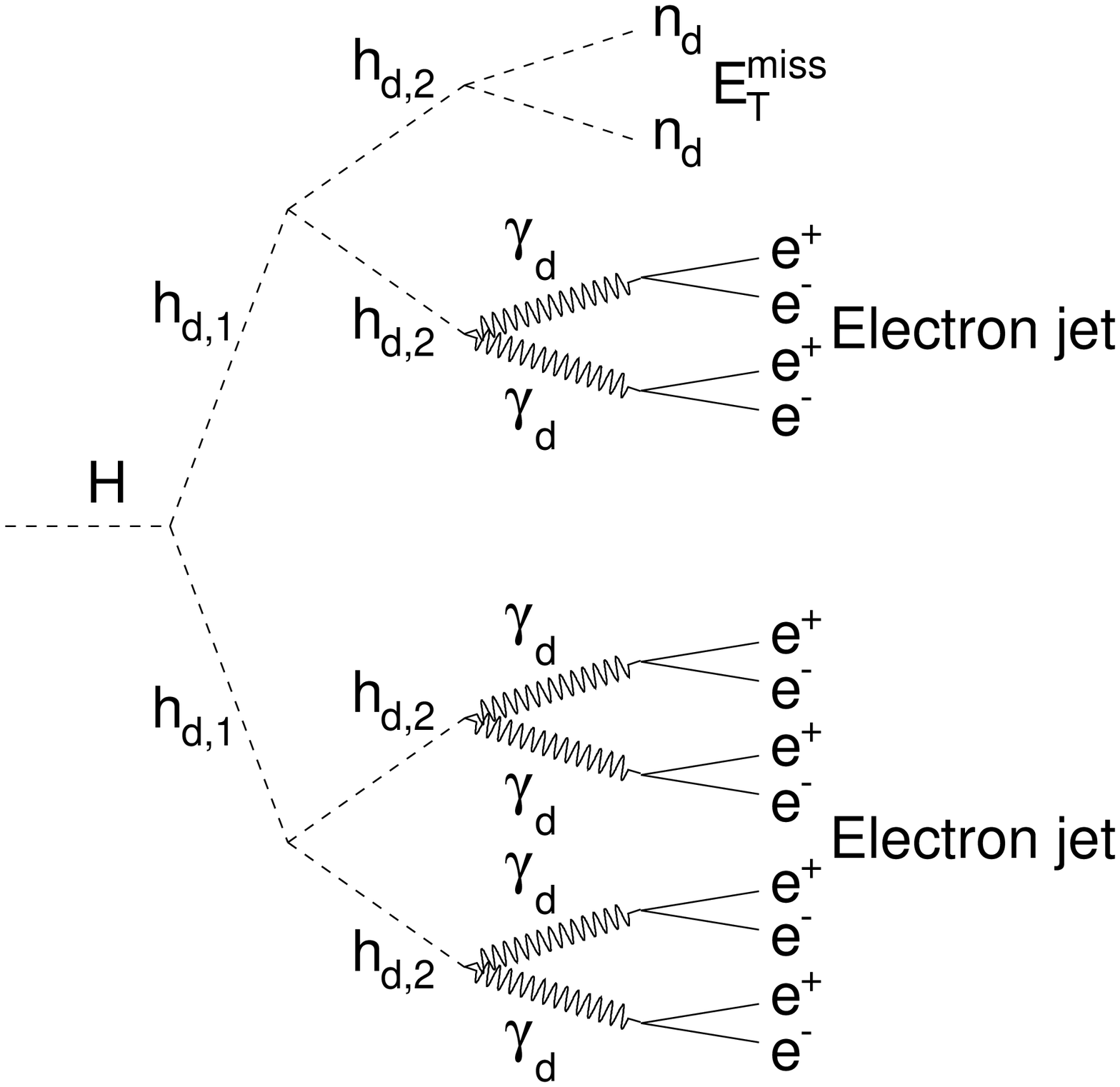}
    \includegraphics[scale=0.28]{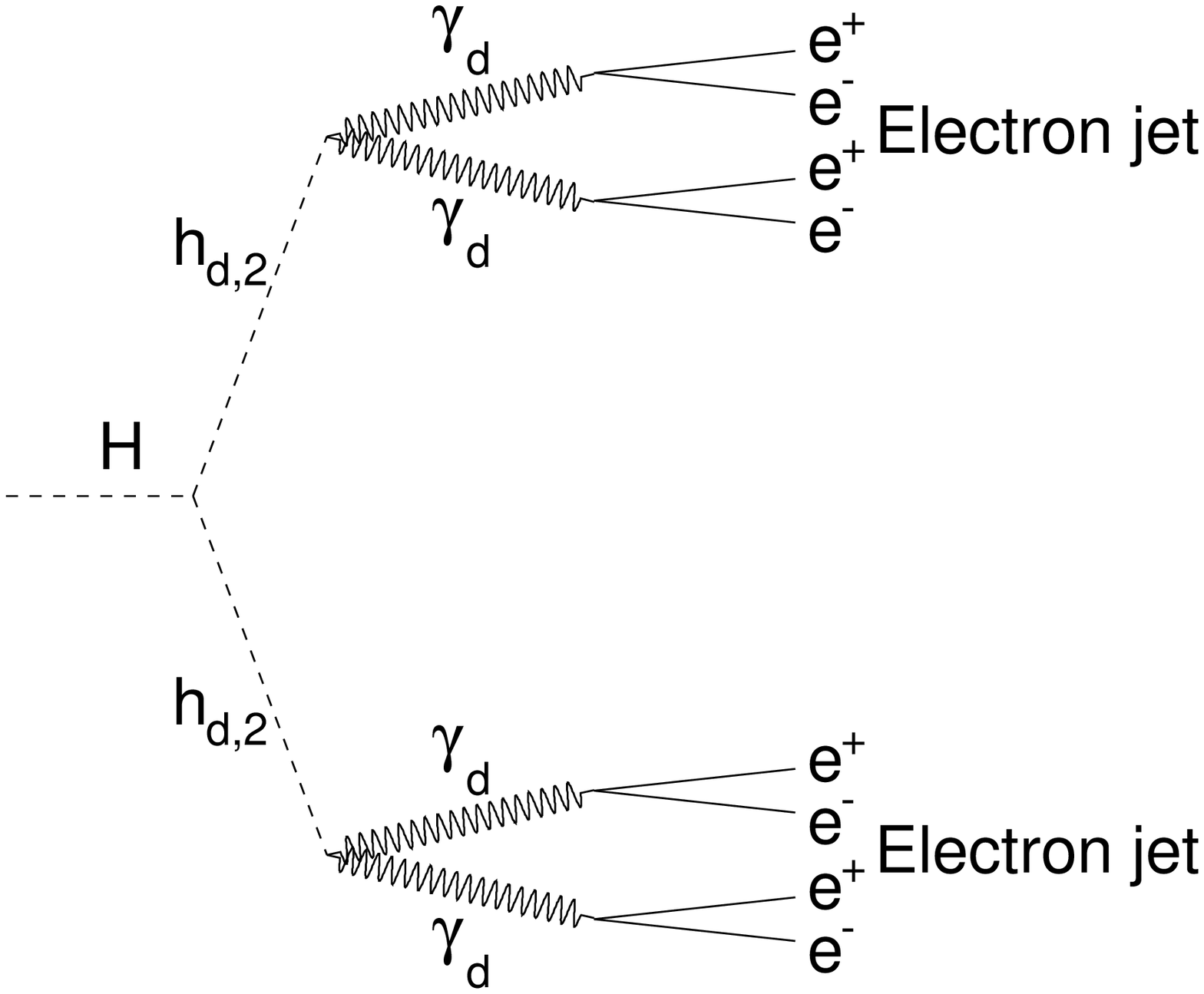}
  \caption{ Diagrams illustrating the Higgs boson decay to hidden-sector particles
in the (left) three-step and (right) two-step models. Each $h_{d,2}$ particle can decay to a pair of dark photons $\gamma_d$ or stable scalars $n_d$, with the corresponding branching ratios given in Table~\ref{tab:hidden_sector_particles}.}
  \label{fig:higgs_production_and_decay}
\end{figure}

In the present analysis two models, discussed in Ref. \cite{discovering_hidden_higgs}, are considered.
These differ in the way the Higgs boson decays, either via a three-step cascade (Fig.~\ref{fig:higgs_production_and_decay} left) or a two-step
cascade (Fig.~\ref{fig:higgs_production_and_decay} right) to hidden-sector particles. In both models the masses of particles in a hidden-sector cascade are taken to be substantially lower than the Higgs boson mass, thus the Higgs boson decay has a two-jet topology.
The models feature a dark photon $\gamma_{d}$ that kinetically mixes with the
SM photon \cite{Holdom198665, Dienes_9610479v2}, a neutral weakly interacting stable scalar $n_d$ 
and two hidden scalars  $h_{d,1}$ and $h_{d,2}$. A value of the kinetic mixing parameter $\epsilon$ larger than $10^{-5}$ implies  dark photons with very short lifetimes; thus the chosen value of $\epsilon=10^{-4}$, recommended in Ref. \cite{discovering_hidden_higgs}, ensures that the decay products are prompt. The dark photon mass must be less than 2 GeV for these models to provide a viable explanation of the results of cosmic-ray
and dark matter direct-detection experiments ~\cite{1538-4357-482-2-L191, PhysRevLett.102.181101, Adriani:2008zr, Chang:2008aa}, which observe an unexpected excess of cosmic electrons and/or positrons, while there is no observed proton excess.
For a dark photon mass below 210 \mev, the dark photons decay exclusively to $e^+e^-$ pairs; Dark photon masses of 100 \mev ~and 200 \mev ~are considered in this analysis. 

\begin{table}[h]
  \begin{center}
    \begin{tabular}{c|r@{}l}
      \hline 
      Parameter & \multicolumn{2}{c}{ } \\
      \hline
      $m_{h_{d,1}}$          &      $10$     &  $\gev$ \\
      $m_{h_{d,2}}$          &      $4$      &  $\gev$ \\
      $m_{n_d}$                     &      $90$    &  $\mev$ \\
      $m_{\gamma_{d}}$  &      $100,200$  &  $\mev$ \\     
      $\epsilon $ & \multicolumn{2}{c}{$10^{-4}$} \\
      \hline
      \multicolumn{3}{c}{three-step model} \\
      \hline
      BR$(h_{d,1}\to h_{d,2} h_{d,2})$       & \multicolumn{2}{c}{$1$} \\
      BR$(h_{d,2}\to \gamma_{d} \gamma_{d})$ & \multicolumn{2}{c}{$0.8$} \\
      BR$(h_{d,2}\to n_d n_d)$  & \multicolumn{2}{c}{$0.2$} \\
      \hline
      \multicolumn{3}{c}{two-step model} \\
      \hline
      BR$(h_{d,2}\to \gamma_{d}\gamma_{d})$  & \multicolumn{2}{c}{$1$} \\
      BR$(h_{d,2}\to n_d n_d)$  & \multicolumn{2}{c}{$0$} \\
       \hline 
    \end{tabular}
  \caption{Parameters of the benchmark hidden-sector models: hidden-sector particle masses,  the $\gamma - \gamma_{d}$ kinetic mixing, and decay branching ratios of $h_{d,1}$ and $h_{d,2}$ in the three-step and two-step models.}
  \label{tab:hidden_sector_particles}
  \end{center}
\end{table}

The signal has a distinct two-jet topology with each electron-jet having a multiplicity of $\geq 4$ electrons per jet, where the electrons are highly collimated. The specific hidden-sector parameters are given in Table \ref{tab:hidden_sector_particles}, and the chosen masses of the Higgs boson are 100 GeV, 125 GeV and 140 GeV. 
The results of the analysis are expected to be robust with respect to the specific choice of the $h_{d,1}$, $h_{d,2}$ and $n_d$ masses as long as these masses are significantly smaller than the Higgs boson mass, i.e. $m_{h_{d,1|2}} \leq$ 10 GeV.  In particular, as long as the $h_{d,1}$ and $h_{d,2}$  scalars are much lighter than the Higgs boson,  the  $h_{d,1}$ and $h_{d,2}$  are boosted, and their decay products are collimated, resulting in two distinct electron-jets. Also in the three-step model, the results are expected to be robust against the explicit choice of the branching ratio of the $h_{d,2}$ particle  into weakly interacting neutral particles, $n_{d}$, as long as this branching ratio is relatively small, i.e. $\mathrm{BR}(h_{d,2}\to n_d n_d) \leq 0.2$. For this range of branching ratios, 
both the $h_{d,1}$ particles decay to visible decay products with greater than 90\% probability; for larger branching ratios, there will be a considerable fraction of events where only one (or neither) $h_{d,1}$ particle decays to visible decay products.

In this analysis, the $\Wboson$ boson produced in association with the
Higgs boson is reconstructed in the $\Wen$ and $\Wmn$ decay modes in
order to achieve a high efficiency for online event selection and a high signal-to-background
ratio. The signal topology is consequently an isolated large transverse momentum lepton accompanied by missing transverse momentum and two or more electron-jets. 

The sensitivity of this analysis is approximately two orders of magnitude greater than that of a previous search for similar signatures which was performed by the CDF collaboration~\cite{PhysRevD.85.092001}.
The direct searches for prompt decays of dark photons into electron or muon pairs, as well as a search for a Higgs boson decaying into displaced muon-jets, were reported in Refs.~\cite{ATLAS-EXOT-2012-16-002, Chatrchyan:1357916,PhysRevLett.105.211802,Aad:2012kw}.

\section{The ATLAS detector}

The ATLAS detector \cite{atlas} is a multi-purpose particle physics detector with forward-backward symmetric cylindrical geometry\footnote{ATLAS uses a right-handed coordinate system with its origin at the nominal $pp$ interaction point at the centre of the detector. The positive {\it x}-axis is defined by the direction from the interaction point to the centre of the LHC ring, with the positive {\it y}-axis pointing upwards, while the beam direction defines the {\it z}-axis. The azimuthal angle $\phi$ is measured around the beam axis with the polar angle $\theta$ is the angle from the {\it z}-axis. The pseudorapidity is defined as $\eta  = - \ln \tan(\theta/2)$. The cone separation is defined as $\Delta R = \sqrt{(\Delta\phi)^2+(\Delta\eta)^2}$. The transverse momentum $\pt$ is defined as the  momentum perpendicular to the beam axis: $\pt = \sqrt{p_x^2+p_y^2}$.}. The inner detector (ID) provides precise reconstruction of tracks with $|\eta| < $ 2.5. It consists of three layers of pixel detectors close to the beam pipe, four layers of silicon microstrip detector modules in the barrel region with pairs of single-sided sensors (SCT) providing eight hits per track at intermediate radii, and a straw-tube transition radiation tracker (TRT) at the outer radii, providing about 35 hits per track (in the range $|\eta| < $ 2.0). The TRT offers substantial discriminating power between electrons and charged hadrons over a wide momentum range (between 0.5 GeV and 100 GeV) via the detection of X-rays produced by transition radiation. 

The ID is surrounded by a thin superconducting solenoid providing a 2\,T axial magnetic field. 
A high-granularity lead/liquid-argon (LAr) sampling calorimeter measures the energy and the
 position of electromagnetic showers with $|\eta| < $  3.2. 
The total thickness of this calorimeter is more than $24 X_0$ in the barrel and above $26 X_0$ in the endcaps. LAr sampling calorimeters, with copper or tungsten as absorber, are used to measure hadronic showers in the endcap (1.5  $<  |\eta| < $ 4.9), while an iron/scintillator tile hadronic calorimeter measures hadronic showers in the central region ($|\eta| < $  1.7). 
The muon spectrometer (MS) surrounds the calorimeters and consists of three large superconducting air-core toroids, each with eight coils, a system of drift tubes and cathode strip chambers for precision tracking  ($|\eta| <$  2.7), and fast tracking chambers for event selection in real time ($|\eta|<$~2.4). 

A three-level trigger system \cite{attrig} selects events to be recorded for offline analysis. The level-1 trigger is implemented in hardware, operating synchronously with the collisions, and uses a subset of detector information to reduce the event rate from 20 MHz to a maximum level-1 output rate of 75 kHz. This is followed by two software-based trigger levels, level-2 and the event filter, which together reduce the recorded event rate to approximately 300 Hz.

\section{Signal and background simulation}
\label{sec:sim}

Simulated data samples are used to estimate the signal acceptance and efficiency, to optimize the signal selection criteria and to cross-check our understanding of the backgrounds. The final background estimate is determined from the data as described in Section~\ref{sec:bkg}.

The signal Monte Carlo (MC) samples are generated with {\scshape MadGraph}  \cite{madgraph_mc} to simulate the Higgs boson production and decay
 to the hidden sector and {\scshape BRIDGE}  \cite{bridge_mc} to simulate the hidden-sector cascades resulting in electron-jets. The output of these two programs is then interfaced to 
{\scshape PYTHIA}  \cite{pythia_mc} for subsequent hadronization and modelling of the underlying event (UE). All leptonic decays modes of the $\Wboson$ boson ($e \nu_e,\mu \nu_\mu,\tau \nu_\tau$) are included in the signal samples.

The most important sources of background are SM $\Wboson/\Zboson$+jets and $\ttbar$ ~processes. A less important source of background comes from pair production of bosons, $\Wboson\Wboson/\Zboson\Zboson/\Wboson\Zboson$, hereafter referred to as diboson production. These processes result in a lepton + jets event topology.  In addition, multi-jet events in some instances can be misidentified as $\Wboson$-bosons.

Samples of simulated $\Wboson(\to \ell\nu)$+jets and $\Zboson(\to \ell\ell)$+jets events ($\ell=e/\mu/\tau$) are generated using {\scshape ALPGEN}   \cite{1126-6708-2003-07-001} with the MLM matching scheme  \cite{springerlink:10.1140/epjc/s10052-007-0490-5} and interfaced to {\scshape HERWIG} \cite{Corcella:2000bw} for parton shower and fragmentation processes and to {\scshape JIMMY} \cite{Butterworth_Forshaw_Seymour_1996} for UE simulation. Simulated samples for $\ttbar$ processes are generated with  {\scshape MC@NLO}  \cite{Frixione:2002ik} interfaced to {\scshape HERWIG} for parton showering. To study the possible dependence on the specific choice of MC generator, alternate $\Wboson/\Zboson$+jets samples are also generated using SHERPA \cite{Gleisberg:2008ta} with an UE modelling according to Ref. \cite{PhysRevD_36_2019}, and alternate top-quark production samples are generated with {\scshape POWHEG} \cite{1126-6708-2007-11-070} interfaced to {\scshape PYTHIA} for hadronization. The diboson processes are generated with {\scshape HERWIG}. Taus are decayed with {\scshape TAUOLA} \cite{davidson_2010} in both signal and background samples.

The event yields for $\Wln$, $\Zll$ ($ell=e/\mu$) and $t\tbar$ ~processes, which give the largest contribution to background, are scaled using the measured production cross sections \cite{wmunu_zll_atlas,ATLAS-CONF-2011-121}. The contributions from $\Wboson(\to\tau\nu)$ and $\Zboson(\to\tau\tau)$, which are minor sources of background, are obtained using next-to-next-to-leading-order cross-section calculations \cite{PhysRevD_74_114017}. The multi-jet background is obtained using normalized data templates \cite{wmunu_zll_atlas},
since the rate with which multi-jet events mimic the combined signature of a prompt charged lepton accompanied by missing transverse momentum is difficult to simulate accurately.

The {\scshape GEANT4} toolkit \cite{1610988} is used for a detailed simulation of the detector response~\cite{Aad:1267853}. The effect of multiple $pp$ interactions per bunch crossing (pile-up) is modelled by overlaying simulated inelastic proton--proton collisions over the original hard-scattering event. 
The simulated events are then passed through the same reconstruction and analysis chain as the data. 

To calibrate the electron energy and to match the resolution of the electron energy and muon momentum observed in data, corrections are applied to electrons in data and electrons and muons in simulated MC samples according to the prescriptions of Refs.~\cite{electron_perf,ATLAS:2011yga}.

\section{Data samples and trigger selection}
The data sample for this analysis was collected by the ATLAS detector in proton--proton collisions at a centre-of-mass energy of 7 TeV in early 2011. The data sample used was required to be recorded during LHC stable-beam conditions when the ATLAS detector components relevant to this analysis were operating within nominal parameters. The total integrated luminosity of the selected data sample is 2.04 $\ifb$ with a 3.7\% uncertainty  \cite{lumi1,lumi2}.

For the $\Wboson \to e \nu$ channel, at least one reconstructed electron trigger object with transverse energy above $22\gev$ in the region of $|\eta| \leq 2.5$ is required. For the $\Wboson \to \mu \nu$ channel,  a muon candidate trigger object in the region of $|\eta| \leq 2.4$ having transverse momentum  above 18~\gev, reconstructed in both the inner detector and muon spectrometer, is required. The muon trigger object must be consistent with having originated from the interaction region.

\section{Event selection and electron-jet reconstruction}

Signal events are required to have exactly one reconstructed $\Wboson$ boson candidate in the $e \nu$ or $\mu \nu$ decay channel and at least two jets identified as electron-jets.

\subsection{ $\Wboson$ boson selection}
\label{sec:wbos}

A $\Wboson$-decay electron candidate  is  required to pass the tight electron selection criteria  \cite{wmunu_zll_atlas,electron_perf} with $p_{\rm T}>25\gev$ and $|\eta|\leq 2.47$. Electrons in the transition region between the barrel and endcap calorimeters  ($1.37 <|\eta|< 1.52$) are rejected. 
A $\Wboson$-decay muon candidate  is required to be identified in both the ID and the MS subsystems and to have $p_{\rm T}>20\gev$ and $|\eta|< 2.4$. To increase the robustness against track mis-reconstruction, the difference between the ID and MS  $p_{\rm T}$ measurements is required to be less than 15 \gev  ~\cite{wmunu_zll_atlas}. 

To reduce background from multi-jet events, electron and muon candidates are required to satisfy an isolation criterion: the sum of the $p_{\rm T}$ of all tracks in a cone of $\Delta R=0.4$ around the electron (muon) divided by the electron (muon)  $p_{\rm T}$  is required to be less than 0.3 (0.2).

$\Wboson$  boson candidates are required to have missing transverse momentum  $\met{}\geq25\gev$ and exactly one isolated electron or muon. Events with two or more isolated same-flavour leptons are rejected, substantially reducing the background from Drell--Yan production.

The lepton candidate from the $\Wboson$ boson decay is required to match the object that satisfies the trigger selection criteria: the distance between the trigger object and the reconstructed $\Wln$  ($\ell=e/\mu$) lepton is required to be $\Delta R<0.1$. 

To reduce the background from cosmic rays, heavy-flavour production and photon conversions, the $\Wboson$ candidate is required to originate from the primary vertex. In events with multiple vertices along the beam axis, the vertex with the largest $\sum{p_{\rm T}^{2}}$, where the sum is over all tracks associated with the vertex, is taken to be the primary vertex of the event. The longitudinal and  transverse impact parameters of the charged-lepton track with respect to the primary vertex must be less than 10 mm and 0.1 mm, respectively.

\begin{figure*}[h]
 \centering
\subfigure[]{
    \includegraphics[scale=0.36]{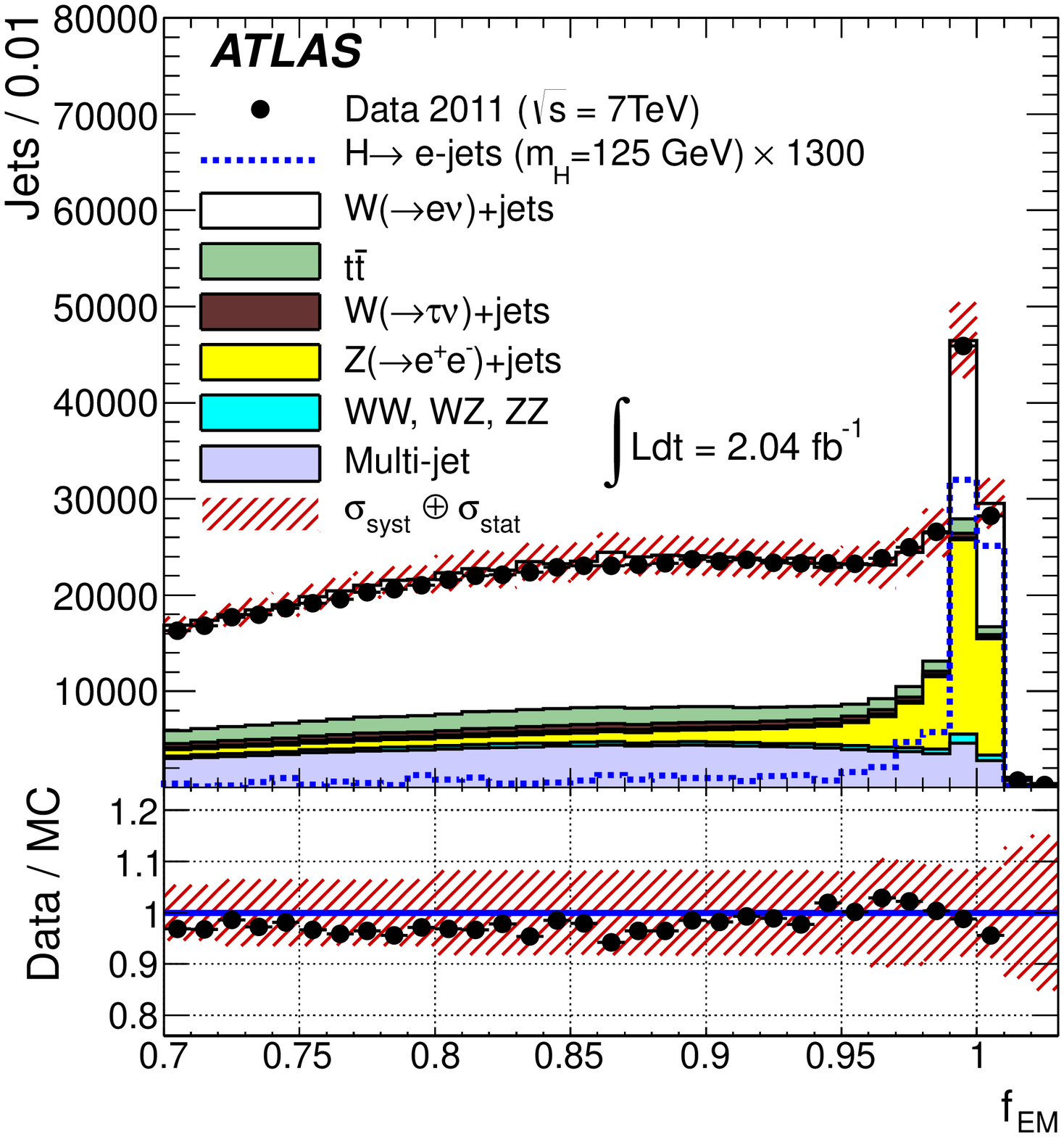}
}
\subfigure[]{
    \includegraphics[scale=0.36]{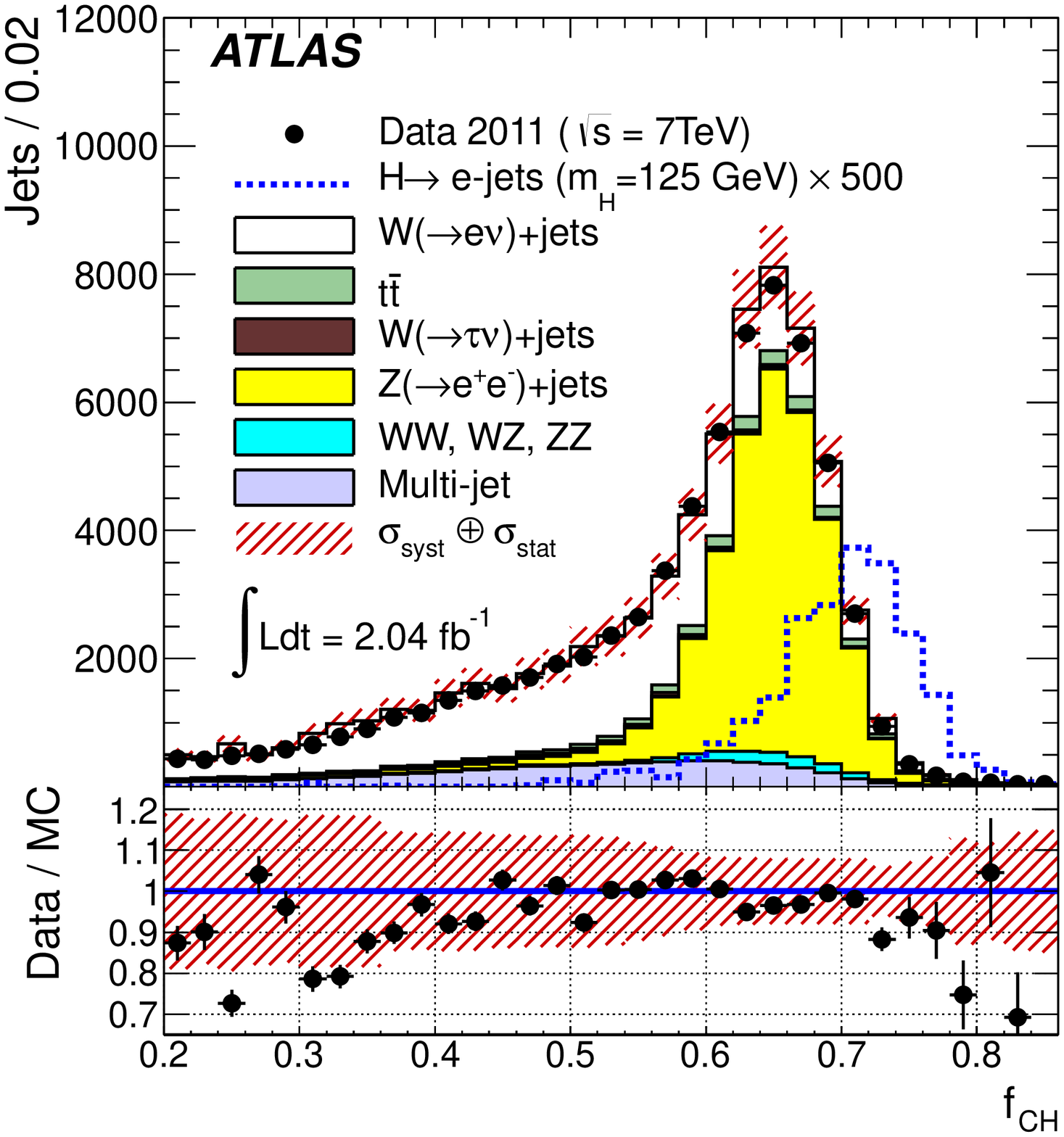}
}
\subfigure[]{
    \includegraphics[scale=0.36]{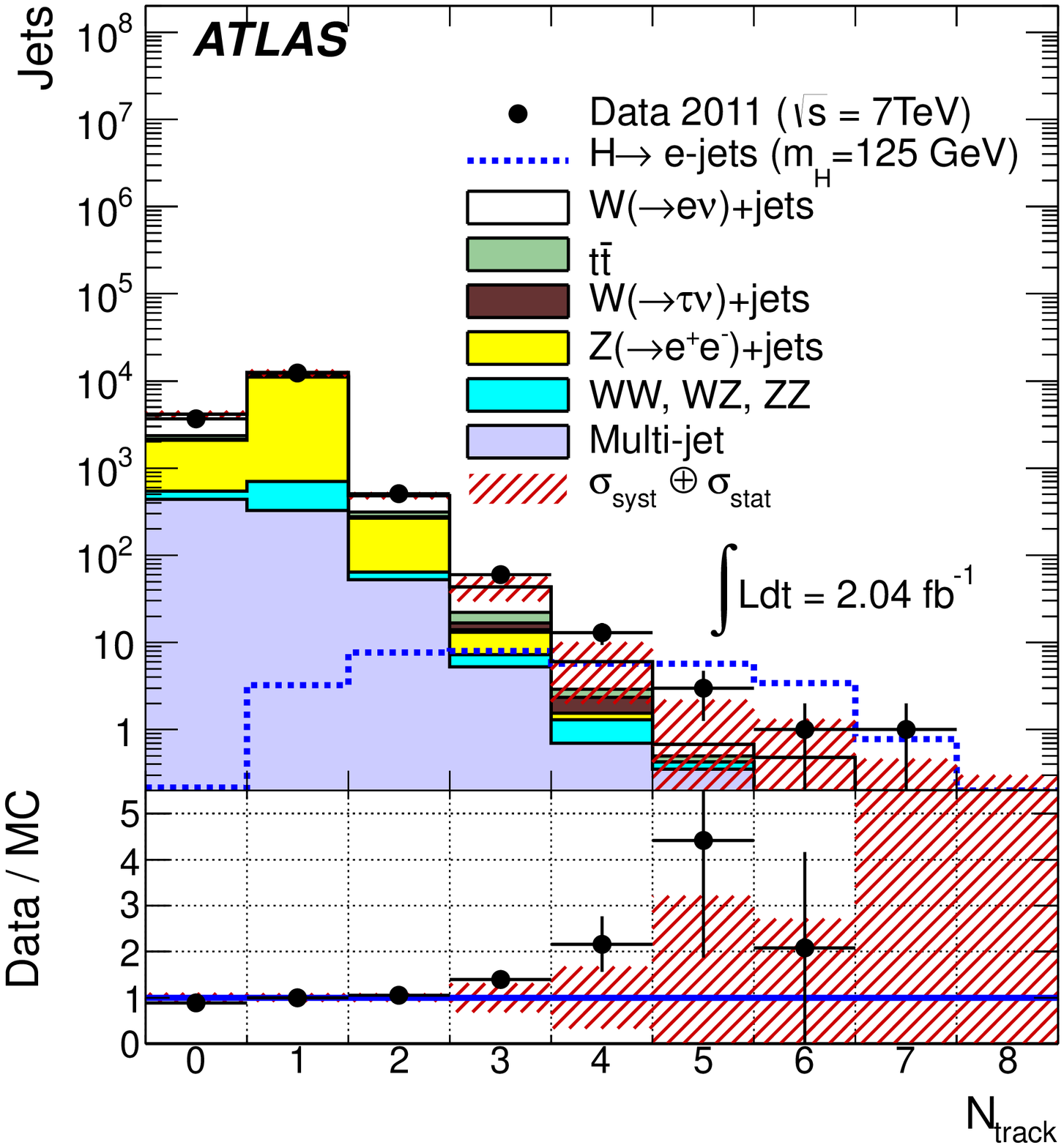}
}
  \caption{Distribution of the jet electromagnetic fraction, $f_{\rm EM}$, after the $\Wen$ selection (a), the jet charged particle fraction, $f_{\rm CH}$, after the $f_{\rm EM}$ selection (b) and the number of associated tracks, $N_{\rm track}$, fulfilling the criteria of Section~\ref{sec:ejet} after the $f_{\rm EM}$ and $f_{\rm CH}$ selection (c). 
 Data are shown by dots with error bars and are compared to the expectation from Standard Model processes, given by stacked histograms of different colors. The signal distributions in
the three-step model of a hidden sector with dark photon mass $m_{\gamma_d}=100$ \mev ~are presented as dashed histograms with arbitrary scale (a and b) and with the nominal scale (c), where nominal scale implies the SM value for $WH$ production cross section and 100\% branching ratio of a Higgs boson decaying to electron-jets.
 The hatched bands represent the quadratic sum of statistical and systematic uncertainties of the Standard Model background prediction described in Section~\ref{sec:syst}.} 
  \label{fig:data_vs_mc_chf_emf} 
\end{figure*}

\subsection{Electron-jet pair selection}
\label{sec:ejet}

In this search, the signature of the Higgs boson decay is
two or more electron-jets. Electron-jet candidates are formed from jets reconstructed in the calorimeters using an  anti-$k_{t}$  jet clustering algorithm  \cite{Cacciari:2008gp}  with a radius parameter $R=0.4$.

The electrons in an electron-jet are too closely collimated to be identified efficiently with the algorithm used to identify electrons from $\Wboson$ boson decays. Instead, electron-jets are identified with three discriminating observables, which are described in detail below: the jet electromagnetic fraction ($f_{\rm EM}$), the jet charged particle fraction ($f_{\rm CH}$) and the fraction of high-threshold hits originating from transition radiation in the TRT ($f_{\rm HT}$).

In electron-jets, the electrons typically deposit all of their energy in the electromagnetic calorimeter, so that the fraction of the jet energy deposited in the electromagnetic calorimeter divided by the total jet energy deposited in both the electromagnetic and hadronic calorimeters, $f_{\rm EM}$, is typically close to unity. The slight degradation of $f_{\rm EM}$ towards lower values is due to the occasional leakage of electromagnetic showers into the hadronic calorimeter, calorimeter noise and electron-jets overlapping with ordinary jets. 
Hadronic jets reaching the calorimeters mainly consist of $\pi^\pm$ and photons from $\pi^0$ decays. Most $\pi^\pm$ deposit a sizable fraction of their energy in the hadronic calorimeter, while photons deposit almost all their energy in the electromagnetic calorimeter. The distribution of $f_{\rm EM}$ is further broadened by fluctuations of the electromagnetic and hadronic showers in the detector. The pedestal corrections for noise in the hadronic calorimeter can sometimes lead to reconstructed energies in the hadronic calorimeter that are less than zero, resulting in a value of $f_{\rm EM}$ slightly higher than unity. The simulation models this situation accurately. The distribution of $f_{\rm EM}$ for hadronic jets peaks around 0.85, with a few percent of these jets having $f_{\rm EM} \geq0.99$. 

Since $f_{\rm EM}$ only provides limited background rejection additional variables are exploited. The quantity $f_{\rm CH}$ is defined as the fraction of the jet energy deposited in calorimeter cells that are associated with tracks within the jet: 
\begin{equation} \label{eq:chf}
f_{\rm CH} = {     \displaystyle\sum_{\rm  ``track-cells" }  E^{\rm cell}    \over        E^{\rm jet}   }.
\end{equation}
A track is associated with a jet if it is within a distance of ${\Delta}R=0.4$ from the jet axis and has $\pt\geq400\mev$. 
The ``track-cells'', i.e. calorimeter cells associated with tracks within the jet, consist of the cells within a cone of $\Delta R =0.2$ around each of the tracks associated with the jet, giving the sum of energy deposits of charged particles within the jet in both the electromagnetic and hadronic calorimeters. 

Signal electron-jets consist exclusively of electrons and should have large $f_{\rm EM}$ and $f_{\rm CH}$.  Hadronic jets with large $f_{\rm EM}$ are expected to contain mostly neutral pions decaying to photons and, therefore, fewer charged tracks and low $f_{\rm CH}$. Photons that convert to electron--positron pairs in the material they traverse before entering the calorimeter increase the value of $f_{\rm CH}$. 

Additional rejection of hadronic jets is achieved by exploiting the identification of electrons using transition radiation. The discriminating quantity is the fraction of TRT hits on a track, $f_{\rm HT}$, that exceed the high discriminator threshold in the read-out electronics of the TRT straws. Detailed studies of this quantity are reported in Refs.~\cite{electron_perf,electron_perf_conf}. This high-threshold setting corresponds to the size of the large energy deposit from transition radiation in the straw-tube gas.
The distribution of $f_{\rm HT}$  for tracks from electrons has a maximum at $f_{\rm HT}$ $\sim$ 0.2, while it peaks at zero and then decreases monotonically for charged hadron tracks. A requirement that $f_{\rm HT} \geq$ 0.08 has an efficiency of over 95\% for electrons in the momentum range relevant to this analysis  and at the same time effectively rejects charged hadrons.
\begin{table*}[ht]
  \begin{center}
    \begin{tabular}{r|l|r|r@{ $\pm$ }l|r@{ $\pm$ }l|l}    
    \hline    
    \multicolumn{2}{c|}{}        & Data      & \multicolumn{2}{c|}{Expected}  & \multicolumn{2}{c|}{Expected }& Eff.\\
    \multicolumn{2}{c|}{}        &           & \multicolumn{2}{c|}{background} & \multicolumn{2}{c|}{signal, $m_{H}$=125 GeV }& \\
     \hline            
	\multicolumn{2}{c|}{} & \multicolumn{6}{c}{$\Wen$ channel}\\ 
  \hline
    \multicolumn{2}{c|}{$\Wen$ selection}  & 4351732   &  4330000  & 250000                   &   46.7 & 1.8$^{\mathrm{(stat)}}$    & 13\%\\
  \cline{1-2}
  \multirow{5}{*}{$\geq$2 jets}  & \pt$\geq$30 GeV,       &  & \multicolumn{2}{c|}{}  &  \multicolumn{2}{c|}{}  &\\
                                 &  $|\eta|\leq2.0$    & 173551 & 183000 & 16000         & 25.6 & 1.4$^{\mathrm{(stat)}}$ & 7.1\%\\  
                                    
    & $f_{\rm EM}\geq$0.99  & 837    & 1070  & 200           &10.8 & 0.9$^{\mathrm{(stat)}}$ & 3.0\% \\   
    & $f_{\rm CH}\geq$0.66  &  39    & 35    &  8            & 6.3 & 0.7$^{\mathrm{(stat)}}$ & 1.7\% \\ 
    & $N_{\rm track} \geq 2$                          &   0    & \multicolumn{2}{c|}{0.10 $^{+~0.11}_{-~0.10}$} & 5.3 & 0.6$^{\mathrm{(stat)}}$ $\pm$ 0.4$^{\mathrm{(syst)}}$  & 1.5\% \\

     \hline            
	\multicolumn{2}{c|}{}  & \multicolumn{6}{c}{$\Wmn$ channel}\\ 
   \hline
    \multicolumn{2}{c|}{$\Wmn$ selection}  & 8870713   &  8620000  & 350000                   &   60.3 & 2.3$^{\mathrm{(stat)}}$  & 17\% \\
    \cline{1-2}    
    \multirow{5}{*}{$\geq$2 jets}  & \pt$\geq$30 GeV,      &  &  \multicolumn{2}{c|}{}   &  \multicolumn{2}{c|}{}  & \\
                                   & $|\eta|\leq2.0$      & 326956 &  353000 & 33000 &  31.5 & 1.7$^{\mathrm{(stat)}}$  & 8.8\%\\
                                   
    & $f_{\rm EM}\geq$0.99  & 1008   &  1240   &  180  &  13.9 & 1.2$^{\mathrm{(stat)}}$ & 3.9\% \\
    & $f_{\rm CH}\geq$0.66  & 45     &   41    &  16   &   7.5 & 0.9$^{\mathrm{(stat)}}$ & 2.1\%\\    
    & $N_{\rm track} \geq 2$                         & 1      &  \multicolumn{2}{c|}{0.11 $^{+~0.13}_{-~0.11}$} &  6.0 & 0.8$^{\mathrm{(stat)}}$ $\pm$ 0.4$^{\mathrm{(syst)}}$ & 1.7\% \\
    \hline    
    \end{tabular} 
  \end{center}
  \caption{The expected number of background and signal events in 2.04~$\ifb$ of data, as well  as the number of events observed in the data, after applying the various signal selection criteria for the $\Wen$ and $\Wmn$ channels. The ``$\geq$2 jets'' in the first column denotes the requirement of two or more jets per event, satisfying the corresponding selection criteria. The signal predictions correspond to the three-step model with  $m_{H}$ = 125 GeV and $m_{\gamma_d}$=100 \mev. The signal efficiencies are the fraction of signal events  satisfying all the selection criteria up to and including that particular criterion. They are given with respect to the signal sample including all three decay modes of the $\Wboson$ boson ($e\nu_e$,~$\mu\nu_{\mu}$,~$\tau\nu_{\tau}$).  The background expectations include statistical and systematic uncertainties, and are determined using the MC method (see Section~\ref{sec:bkg}). The statistical uncertainty shown for the signal is due to Monte Carlo statistics; the systematic uncertainty is only given for the final event selection and its detailed composition is given in Table~\ref{tab:all_systematics}.}
\label{tab:cutflow}
\end{table*}
Exploiting the selection criteria described above, a jet is classified as an electron-jet if it satisfies the following requirements:
\begin{itemize}
\item  jet pseudorapidity $|\eta|\leq2.0$,
\item  jet transverse momentum $\pt\geq30\gev$,
\item  jet electromagnetic fraction $f_{\rm EM}\geq0.99$,
\item  jet charged particle fraction $f_{\rm CH}\geq0.66$ and 
\item  number of tracks associated with the jet $N_{\rm track} \geq 2$, where the tracks must satisfy the following criteria: 
\begin{itemize}
\item  track pseudorapidity $|\eta|\leq 2.0$,
\item  track transverse momentum $\pt\geq 5 \gev$,
\item  number of hits in the pixel detector $N_{\rm PIX}\geq2$,
\item  total number of pixel and SCT hits $N_{\rm PIX}+N_{\rm SCT}\geq7$, 
\item  fraction of high-threshold TRT hits $f_{\rm HT}\geq 0.08$. 
\end{itemize}
\end{itemize}

Good agreement is observed between data and MC simulation in the $f_{\rm EM}$, $f_{\rm CH}$ and track-related distributions at the different stages of selection, as can be seen in  Fig. \ref{fig:data_vs_mc_chf_emf}. The number of events observed in the data and the yields expected for the background and the signal as the selection criteria are applied are shown in Table \ref{tab:cutflow}. The background yield given here is determined by MC.

\section{Background estimation }
\label{sec:bkg}

The dominant background in this search is due to the associated production of a $\Wboson$ boson with hadronic jets which mimic the electron-jet signature. Detailed MC studies of the background contamination from hadronic jets faking electron-jets have shown that the high electron content in those jets originates either from final-state photon radiation or from $\pi^0$ decays with subsequent photon conversions in the material of the detector.
A background prediction from MC simulation would depend on the modelling of final-state photon radiation and parton showering and hadronization, which would introduce large uncertainties in the background rate. Instead, the background contamination in the signal region is estimated from the data using a simplified matrix method \cite{toppaper} 
which is completely data-driven. 
Two alternative background estimates were tried and found to be consistent with the matrix method result. One of these estimates -- referred to as the ABCD method below -- is based on data; the second estimate is based on Monte Carlo simulation.

\begin{table*}[t]
\begin{center}
\begin{tabular}{c@{ }c|c}
    \hline
    \multicolumn{2}{c|}{Method of background estimation} & Estimated background yield \\
    \hline
    Matrix & (baseline)        &  0.41 $\pm$ 0.29 (stat) $\pm$  0.12 (syst) \\
    ABCD   & (cross-check)     &  0.46 $\pm$ 0.32 (stat) $\pm$  0.10 (syst) \\
    MC     & (cross-check)     &  0.21 $\pm$ 0.05 (stat) $^{+}_{-}$  $^{0.23}_{0.21}$~~~(syst) \\
    \hline 
\end{tabular}  
\end{center}
\caption{Estimated number of background events after the final selection, including statistical and systematic uncertainties, from three different methods.  The matrix method is used 
for the background estimate while the ABCD and MC methods provide a cross-check of the matrix method.}
\label{tab:all_background_summary}
\end{table*}

In the matrix method,  one defines a ``loose'' electron-jet selection criterion by relaxing the minimum track $p_{\rm T}$ requirement from 5$\gev$ to 2$\gev$. The fraction $f$ is the ratio of the number of jets passing the nominal signal selection  $N_{T}$ to that passing the loose selection $N_{L}$:
\begin{equation}
f=\frac{N_{T}}{N_{L}}.
\end{equation}
 The number of fake electron-jet background events passing the nominal selection criteria for two electron-jet candidates and entering the signal region is therefore:
\begin{equation}
n_{\mathrm{bkgd}}= {f^2} n_{\mathrm{fake}},
\label{eq:f1}
\end{equation}
where $n_{\mathrm{fake}}$ is the number of background events passing the loose criterion for both electron-jet candidates.   
On the other hand, the number of fake electron-jet events in which neither electron-jet candidate passes the nominal selection criterion is 
\begin{equation}
n_{\mathrm{loose}}= {(1-f)^2} n_{\mathrm{fake}}.
\label{eq:f2}
\end{equation}

Combining Eqs. (\ref{eq:f1}) and (\ref{eq:f2}) yields
\begin{equation}
n_{\mathrm{bkgd}}=\frac{f^2}{(1-f)^2}n_{\mathrm{loose}}=\bar{f}^{2}n_{\mathrm{loose}},
\end{equation}
where $\bar{f}$ is referred to below as the fake factor. In this way the number of background events is derived directly from the data events failing to pass the nominal criteria  for both electron-jet candidates ($ n_{\mathrm{loose}}$), where the signal contamination has been checked to be small. 

The fake factor $\bar{f}$ is measured from background-enriched data samples where the signal contamination is checked to be completely negligible. The first sample is obtained by reversing the $\Wboson$-candidate electron or muon selection criteria to select a sample of multi-jet background with kinematic characteristics similar to those of the signal sample. A fake factor is obtained from the jets in this sample. In the second sample,  the fake factor is determined from a sample of jets that originate from electrons in $\Zboson\to e^+e^-$ decays. The tight selection criteria and the lepton isolation criteria are applied to one leg of the $\Zboson$ boson candidate, and the invariant mass of this electron and the candidate jet, $m_{e,\mathrm{jet}}$, is required to fall in the range $80~\mathrm{GeV} < m_{e,\mathrm{jet}} < 100~\mathrm{GeV}$. The two fake factors are found to be consistent within statistical uncertainties: $0.44 \pm 0.02$ (stat) and $0.47 \pm 0.03$ (stat), respectively. The first value is used in the analysis and the difference between these two estimates is taken as a systematic uncertainty in the fake factor. The resulting value is $\bar{f}=0.44 \pm 0.04$, where the uncertainty is the sum of the statistical and systematic uncertainties added in quadrature. 

In the ABCD method used to cross-check the results, events are assigned to one of four regions according to whether or not the jets meet the $f_{\rm EM}$ and the track-quality conditions of the electron-jet classification. These two conditions are chosen because they are less correlated than other selection variables that could have been used. The background yield in the signal region is thus given by
\begin{equation}
n_{\mathrm{bkgd}} = N_{A}^{\mathrm{predicted}} = \frac{N_{B}\cdot N_{C}}{N_{D}} \cdot \mathrm{c_{MC}}~\mathrm{,}
\end{equation}
where $\mathrm{c_{MC}}=0.36$ is the correction factor determined from MC simulation that corrects for the effect of the correlations between the two selection criteria. $N_{i}$ is the number of events observed in region $i$.

The regions are defined as:
\begin{list}{}{}
\item {(A)  Signal region:} Two jets with $f_{\rm EM}\geq 0.99$ are required; for both jets $N_{\rm track} \geq 2$, i.e. the number of tracks associated with the candidate jets and fulfilling the requirements of the electron-jet selection must be greater than two (see Section~\ref{sec:ejet}). 
\item {(B)  Anti-track quality region:}   At least one jet must fail the associated track requirements of the electron-jet selection. Each of the two candidate jets must have two associated tracks separated by $\Delta R<0.1$, both tracks satisfying looser requirements $\pt\geq 2 \gev$ and $f_{\rm HT} > 0$.  At least one of these tracks must fail the requirement $f_{\rm HT} > 0.08$.
\item {(C)  Anti-$f_{\rm EM}$ region:} At least one jet must fail the condition $f_{\rm EM}\geq 0.99$.
\item {(D)  Anti-track quality and anti-$f_{\rm EM}$ region:} Both conditions ``B'' and ``C'' are fulfilled.
\end{list}

In the second cross-check of the matrix method, the background prediction is obtained by using data templates and simulated samples with the appropriate cross sections scaled by the measured integrated luminosity, as described in Section~\ref{sec:sim}.  

The resulting background yields, together with the evaluated statistical and systematic uncertainties, are given in Table \ref{tab:all_background_summary}. The estimates from the different background evaluation methods agree well within the uncertainties. 

\section{Systematic uncertainties}
\label{sec:syst}
The systematic uncertainties considered for the signal are given in Table~\ref{tab:all_systematics} and described in detail below:
\begin{table}
  \begin{center}
    \begin{tabular}{c@{ }l|r@{\% }r} 
    \hline    
    \multicolumn{2}{c}{Systematic Source}  & \multicolumn{2}{c}{Systematic Uncertainty} \\
    \hline
    (i) & Monte Carlo statistics  		&~~~~~~~~~~~ 13&  \\    
    (ii) & Luminosity                          	&  3.7& \\
    (iii) & $\sigma\times\mathrm{BR}$           		& $^{+3.7}_{-4.3}$& \\ 
     (iv) & Electron efficiency 	                & 5&   \\   
    (iv) & Muon efficiency         		        & 3& \\
    (v) & $f_{\rm EM}$ modelling   		& 3&   \\
    (v) & $f_{\rm CH}$ modelling    		& 0.1& \\
    (vi) & $f_{\rm HT}$ modelling     		& 1&   \\
    (vii) & Electron energy scale   			& 0.5& \\    
    (vii) & Electron energy resolution         	& 0.2& \\
    (vii) & Muon momentum resolution           	& 0.5& \\      
    (viii) & Pile-up 		     	  		& $<$0.1&  \\    
    \hline
    \multicolumn{2}{c}{Total} 				&  15& \\
    \hline
    \end{tabular}
  \end{center}
  \caption{Systematic uncertainties for the signal. The numbers in parentheses refer to the descriptions in the numbered list in the text. All uncertainties are applied to the combination of $\Wen$ and $\Wmn$ channels; the only exceptions are the specific electron and muon uncertainties in items (iv) and (vii), and are applied separately. The total uncertainty is conservatively rounded and is given for the combination of channels.} 
\label{tab:all_systematics}
\end{table}

\begin{enumerate}

\item {\bf Monte Carlo statistics:} The uncertainty due to the limited number of Monte Carlo signal events is 13\%.  

\item{\bf Luminosity}: The uncertainty in the integrated luminosity is determined to be 3.7\% \cite{lumi1,lumi2}.

\item {\bf Signal cross sections:} The uncertainty of the SM $WH$ production cross section at Higgs mass $m_{H}=$ 125 GeV is $_{-4.3\%}^{+3.7\%}$~\cite{Higgs_cross-sections}. For 100 GeV and 140 GeV Higgs mass the corresponding uncertainties are $\pm$4\%~\cite{Higgs_cross-sections}. 

\item {\bf Electron and muon efficiency:} The combined uncertainty on the efficiency of the lepton trigger, identification and isolation as well as transverse impact parameter requirements is found to be 5\% for electrons and 3\% for muons. The uncertainties were derived using data-driven methods \cite{electron_perf,ATLAS:2011yga}. 

\item {\bf Jet electromagnetic and charged particle fractions ($f_{\rm EM}$ and $f_{\rm CH}$):} The uncertainty due to possible mismodelling of these parameters, which impacts the signal acceptance, are studied by comparing the measured $f_{\rm EM}$ and $f_{\rm CH}$ line shape for jets, which are matched to electron from $\Wboson$-decay, to the one predicted by the simulation. They are found to be 3\% and 0.1\% for $f_{\rm EM}$ and $f_{\rm CH}$, respectively.

\item {\bf Fraction of high-threshold hits in the TRT ($f_{\rm HT}$):} 
Mismodelling of the $f_{\rm HT}$ distribution in the simulation has been previously studied~\cite{electron_perf,electron_perf_conf}.
The impact of this mismodelling on the signal efficiency was checked using the data samples, enriched with highly collimated pairs of electron tracks, and is found to be less than 1\%.

\item {\bf Electron and muon energy/momentum scale and resolution:} 
These uncertainties are evaluated by varying the corresponding correction factors, described in Section~\ref{sec:wbos}, within their systematic uncertainties. This results in the corresponding uncertainties  of 0.2\%, 0.5\% and 0.5\% for electron/muon energy resolution and electron energy scale respectively.

\item {\bf Pile-up impact:} The effect of additional inelastic collisions overlapping with the primary hard scatter (pile-up) on the signal efficiency has been evaluated using simulated signal samples and found to be negligible.

\end{enumerate}

The systematic uncertainties on the background determinations are estimated in the following ways: 
\begin{itemize}
\item {\bf Matrix method:} The uncertainty is assessed by varying the fake factor within its uncertainty and is summed in quadrature with the statistical uncertainty on the number of events observed in the loose region. An uncertainty due to possible signal contamination is also taken into account. These result in an 80\% uncertainty in the background yield.

\item  {\bf ABCD method:} The uncertainty is assessed  by employing different region selections in the ABCD method and the difference between yields is treated as a systematic uncertainty. The uncertainty due to limited statistics in region B is also considered. The resulting uncertainty in the background yield is 70\% with this method.

\item {\bf MC method}: The largest contributions to the uncertainty on the  background yield  are 
the systematic uncertainty on the MC-based prediction on the probability of two or more jets to be incorrectly identified as electron-jets (100\%), modelling of $f_{\rm HT}$ (50\%), $f_{\rm CH}$ (10\%), $f_{\rm EM}$ (10\%), the choice of MC generator (10\%) and uncertainties on the cross section for multi-jet processes (10\%). The assigned theoretical uncertainties on the cross sections are 4\% ($\Wln$, $\Zll$), 5\% ($W\to\tau\nu$, $Z\to\tau\tau$, $WZ/ZZ$) and 7\% ($\ttbar$, $WW$), respectively \cite{wmunu_zll_atlas,ATLAS-CONF-2011-121}. Limited MC sample sizes contributes a 5\% uncertainty. The total uncertainty of background estimation with MC method is 110\%. 
\end{itemize}

\section{Results and interpretation}

The observed and predicted event yields after the final selection are shown in Table \ref{tab:event_yields_summary}. The event yield in the signal region is consistent with the background-only hypothesis, with one data event passing the final selection in the $\Wboson \to \mu \nu$ channel and no data events passing the final selection in the $\Wboson \to e \nu$  channel.

\begin{table*}[htb]
 \begin{center}
    \begin{tabular}{c|r@{.}l|r@{.}l|r@{.}l|r@{.}l}
    \hline
     Signal &  \multicolumn{4}{c|}{three-step model} & \multicolumn{4}{c}{two-step model} \\
    \hline
     $m_{H}$ (\gev) & \multicolumn{2}{c|}{$m_{\gamma_{d}}$ = 100 \mev}  & \multicolumn{2}{c|}{$m_{\gamma_{d}}$ = 200 \mev}  & \multicolumn{2}{c|}{$m_{\gamma_{d}}$ = 100 \mev}  & \multicolumn{2}{c}{$m_{\gamma_{d}}$ = 200 \mev} \\  
\hline
    100 &  14&3 $\pm$ 1.7 $\pm$ 0.8 & 12&4 $\pm$ 1.6 $\pm$ 0.7 & 22&6 $\pm$ 2.1 $\pm$ 1.2 & 23&5 $\pm$ 2.1 $\pm$ 1.2 \\
    125 &  11&3 $\pm$ 1.0 $\pm$ 0.6 & 10&7 $\pm$ 1.1 $\pm$ 0.6 & 16&2 $\pm$ 1.2 $\pm$ 0.9 & 18&1 $\pm$ 1.4 $\pm$ 1.0 \\
    140 &  9&6 $\pm$ 0.8 $\pm$ 0.5 & 9&0 $\pm$ 0.8 $\pm$ 0.4 & 13&7 $\pm$ 0.9 $\pm$ 0.8 & 13&9 $\pm$ 0.9 $\pm$ 0.8 \\
\hline
    Background   & \multicolumn{8}{c}{0.41 $\pm$ 0.29 $\pm$ 0.12 } \\
 \hline
    Data       &       \multicolumn{8}{c}{1}          \\
    
    \hline
  \end{tabular}
  \end{center}
  \caption{Numbers of expected and observed events after final selection. Expected signal yields are provided for both the three-step and two-step models with dark photon masses of 100 and 200 \mev. Statistical (first) and systematic (second) uncertainties are presented separately. The results are given for the combination of the $\Wen$ and $\Wmn$ channels. One candidate event is observed in the data in the $\Wmn$ channel.}
\label{tab:event_yields_summary}
\end{table*}

\begin{table}[htb]
  \begin{center}
  \begin{tabular}{c|c|c|c}
  \hline
Model & $m_{\gamma_{d}}$ (\mev)  &  Observed & Expected \\ 
  \hline
  three-step & 100  &  0.39 &  0.37 \\
  three-step & 200  &  0.45 &  0.44 \\
  two-step & 100    &  0.29 &  0.28 \\
  two-step & 200    &  0.24 &  0.24 \\
  \hline
  \end{tabular}
  \end{center}
  \caption{The 95\% CL upper limits on the signal strength, $\sigma(WH) \times \mathrm{BR}(H\to e$-jets$)/\sigma_{\mathrm{SM}}(WH)$,
for a Higgs boson mass of 125 GeV for different choices of the hidden-sector model parameters. Here, $\sigma(WH)$ is the $WH$ production cross section times the sum of the branching ratios for the $\Wboson$ boson decaying to leptons ($e\nu_e$,~$\mu\nu_\mu$,~$\tau\nu_\tau$), $\sigma_{\mathrm{SM}}(WH)$ is the corresponding SM expectation  and $\mathrm{BR}(H\to e$-jets$)$ is the branching ratio for Higgs boson decays to electron-jets.}
\label{tab:higgs_exclusion_numbers}
\end{table}

Consequently one estimates a 95\% confidence limit on the signal strength,   $\sigma(WH)\times\mathrm{BR}(H\to e$-jets$)/\sigma_{\mathrm{SM}}(WH)$, where $\sigma(WH)$ denotes the $WH$ production cross section times the sum of the branching ratios for the $\Wboson$ boson decaying to leptons ($e\nu_e$,~$\mu\nu_\mu$,~$\tau\nu_\tau$) and $\sigma_{\mathrm{SM}}(WH)$ is the corresponding SM expectation \cite{Higgs_cross-sections} for this quantity ($\sigma_{\mathrm{SM}}(WH)=223^{+8}_{-10}~\mathrm{fb}$ for the Higgs boson mass of 125 GeV). $\mathrm{BR}(H\to e$-jets$)$ denotes the branching ratio for Higgs boson decays to electron-jets. The results are presented in Fig.~\ref{fig:higgs_mass_exclusion} and Table~\ref{tab:higgs_exclusion_numbers} for both the three-step and two-step models.  Only one limit is presented in Fig.~\ref{fig:higgs_mass_exclusion}  for each of the two-step and three-step models, because  the results are compatible within the statistical uncertainty of the signal for both dark photon masses $m_{\gamma_{d}} = 100 \mev$ and $200 \mev$.   Limits are derived using the CLs technique \cite{0954-3899-28-10-313}. The likelihoods are given by the Poisson distribution for the total number of events in the signal region and are calculated using the number of expected and observed events, whereby the results of the electron and muon channels are summed and enter the likelihood function together. The corresponding signal and background systematic uncertainties are incorporated into the likelihoods as nuisance parameters with Gamma probability density functions~\cite{Cranmer:1456844}. 
Assuming that the $\Wboson H$ cross section has the SM value, for the specific
set of hidden-sector parameters chosen here, the analysis excludes Higgs boson branching ratios to electron-jets between 24\% and 45\% for $m_H=125$ GeV at 95\% confidence level.

\begin{figure*}[htb]
  \centering  
    \includegraphics[scale=0.37]{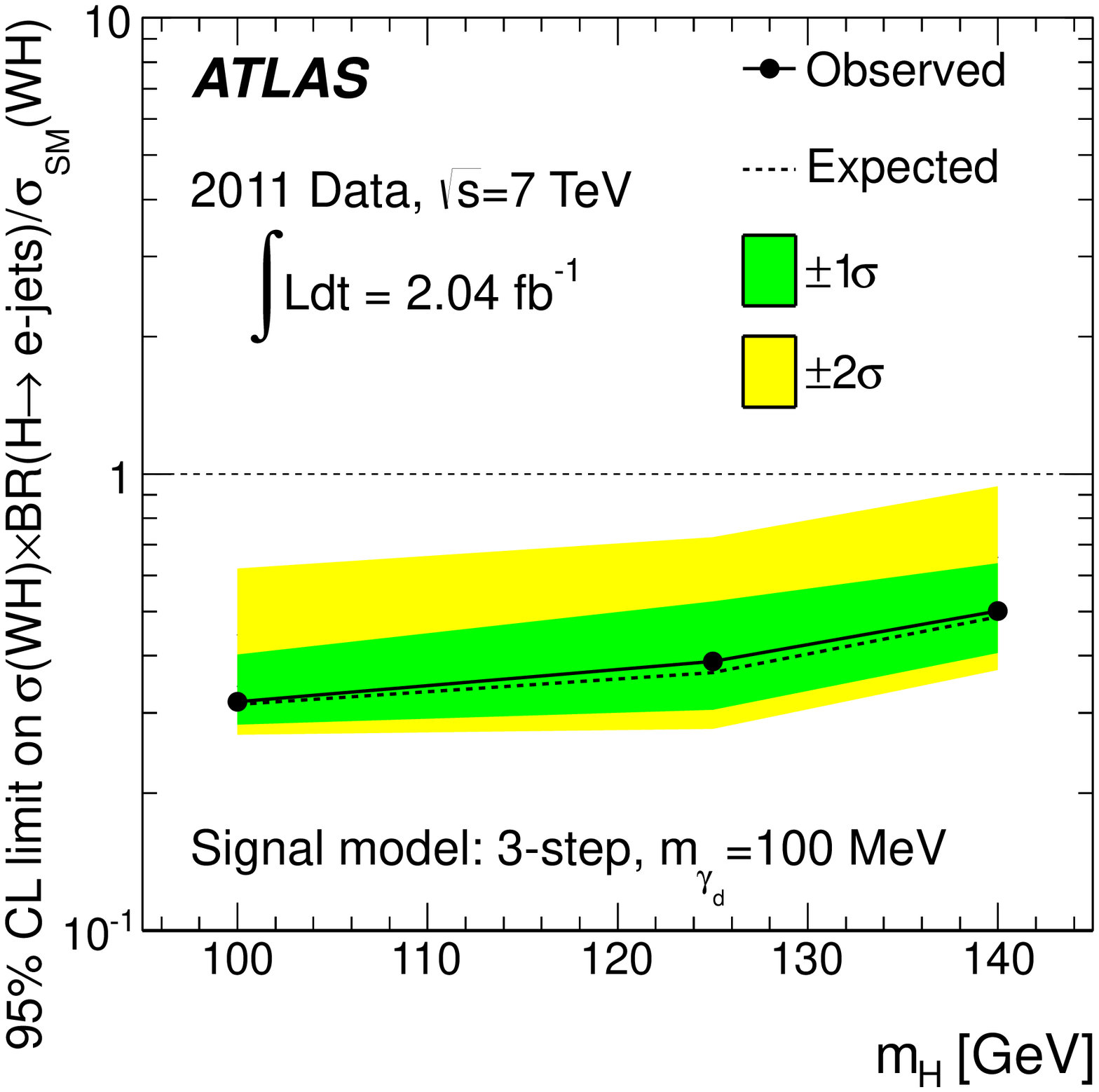}
    \includegraphics[scale=0.37]{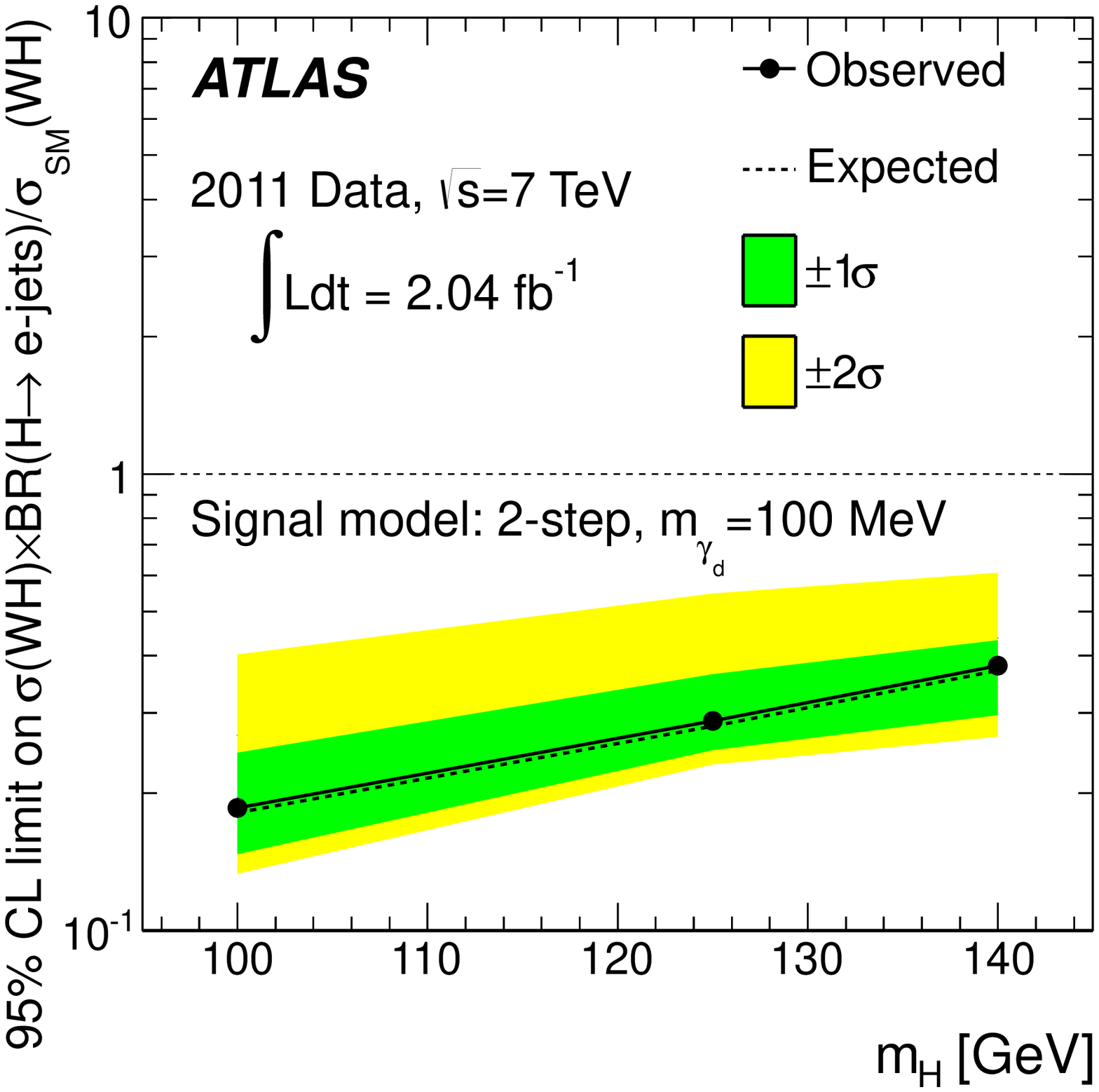}
  \caption{Observed and expected 95\% CL upper limits on the signal strength,
 $\sigma(WH) \times \mathrm{BR}(H\to e$-jets$)/\sigma_{\mathrm{SM}}(WH)$, as a function of the Higgs boson mass for the (left) three-step and (right) two-step models of a hidden sector with a dark photon mass $m_{\gamma_d}=100$ \mev.
The dark (light) shaded band contains 68\% (95\%) of the outcomes of pseudo-experiments generated under the background-only hypothesis.} 
 \label{fig:higgs_mass_exclusion}
\end{figure*}

\section{Conclusions}

A search is presented for a light Higgs boson decaying to a light hidden sector and subsequently into highly collimated jets of electrons, which are expected to be seen in the detector as distinct objects called ``electron-jets''. The analysis has been performed using 2.04 $\ifb$ of proton--proton collision data at $\sqrt{s}$=7 TeV, collected with the ATLAS detector at the LHC in 2011.

The search is performed in the $\Wboson H$ production mode with the choice of hidden-sector parameter space resulting in Higgs boson decaying into prompt electron-jets.
The electron-jet identification method presented here provides good discrimination against background sources and avoids sensitivity to the detailed topology of the electrons within the electron-jet.

The observed data are consistent with the Standard Model background hypothesis. Consequently,  95\%  confidence level limits are set on the $\Wboson H$ production cross section times the branching ratio into electron-jets, assuming the two benchmark models of a hidden sector and the condition of a dark photon mass below 210 \mev.

\section{Acknowledgements}





We thank CERN for the very successful operation of the LHC, as well as the
support staff from our institutions without whom ATLAS could not be
operated efficiently.

We acknowledge the support of ANPCyT, Argentina; YerPhI, Armenia; ARC,
Australia; BMWF and FWF, Austria; ANAS, Azerbaijan; SSTC, Belarus; CNPq and FAPESP,
Brazil; NSERC, NRC and CFI, Canada; CERN; CONICYT, Chile; CAS, MOST and NSFC,
China; COLCIENCIAS, Colombia; MSMT CR, MPO CR and VSC CR, Czech Republic;
DNRF, DNSRC and Lundbeck Foundation, Denmark; EPLANET, ERC and NSRF, European Union;
IN2P3-CNRS, CEA-DSM/IRFU, France; GNSF, Georgia; BMBF, DFG, HGF, MPG and AvH
Foundation, Germany; GSRT and NSRF, Greece; ISF, MINERVA, GIF, DIP and Benoziyo Center,
Israel; INFN, Italy; MEXT and JSPS, Japan; CNRST, Morocco; FOM and NWO,
Netherlands; BRF and RCN, Norway; MNiSW, Poland; GRICES and FCT, Portugal; MERYS
(MECTS), Romania; MES of Russia and ROSATOM, Russian Federation; JINR; MSTD,
Serbia; MSSR, Slovakia; ARRS and MVZT, Slovenia; DST/NRF, South Africa;
MICINN, Spain; SRC and Wallenberg Foundation, Sweden; SER, SNSF and Cantons of
Bern and Geneva, Switzerland; NSC, Taiwan; TAEK, Turkey; STFC, the Royal
Society and Leverhulme Trust, United Kingdom; DOE and NSF, United States of
America.

The crucial computing support from all WLCG partners is acknowledged
gratefully, in particular from CERN and the ATLAS Tier-1 facilities at
TRIUMF (Canada), NDGF (Denmark, Norway, Sweden), CC-IN2P3 (France),
KIT/GridKA (Germany), INFN-CNAF (Italy), NL-T1 (Netherlands), PIC (Spain),
ASGC (Taiwan), RAL (UK) and BNL (USA) and in the Tier-2 facilities
worldwide.

\section*{References}
\bibliography{WplusElectronJets}

\onecolumn 
\clearpage
\begin{flushleft}
{\Large The ATLAS Collaboration}

\bigskip

G.~Aad$^{\rm 48}$,
T.~Abajyan$^{\rm 21}$,
B.~Abbott$^{\rm 111}$,
J.~Abdallah$^{\rm 12}$,
S.~Abdel~Khalek$^{\rm 115}$,
A.A.~Abdelalim$^{\rm 49}$,
O.~Abdinov$^{\rm 11}$,
R.~Aben$^{\rm 105}$,
B.~Abi$^{\rm 112}$,
M.~Abolins$^{\rm 88}$
% ...
\bigskip
\\
% ...
$^{11}$ Institute of Physics, Azerbaijan Academy of Sciences, Baku, Azerbaijan\\
$^{12}$ Institut de F{\'\i}sica d'Altes Energies and Departament de F{\'\i}sica de la Universitat Aut{\`o}noma de Barcelona and ICREA, Barcelona, Spain\\
% ...
$^{21}$ Physikalisches Institut, University of Bonn, Bonn, Germany\\
% ...
$^{48}$ Fakult{\"a}t f{\"u}r Mathematik und Physik, Albert-Ludwigs-Universit{\"a}t, Freiburg, Germany\\
$^{49}$ Section de Physique, Universit{\'e} de Gen{\`e}ve, Geneva, Switzerland\\
% ...
$^{88}$ Department of Physics and Astronomy, Michigan State University, East Lansing MI, United States of America\\
% ...
$^{105}$ Nikhef National Institute for Subatomic Physics and University of Amsterdam, Amsterdam, Netherlands\\
% ...
$^{111}$ Homer L. Dodge Department of Physics and Astronomy, University of Oklahoma, Norman OK, United States of America\\
$^{112}$ Department of Physics, Oklahoma State University, Stillwater OK, United States of America\\
% ...
$^{115}$ LAL, Universit{\'e} Paris-Sud and CNRS/IN2P3, Orsay, France\\
% ...

\end{flushleft}


\end{document}
%